# Creating, Automating, and Assessing Online Homework in Introductory Statistics and Mathematics Classes

Karen Santoro[1] and Roger Bilisoly[1]

[1]Department of Mathematical Sciences, Central Connecticut State University,
1615 Stanley St, New Britain, CT 06050-4010

**Abstract**
Although textbook publishers offer course management systems, they do so to promote brand loyalty, and while an open source tool such as WeBWorK is promising, it requires administrative and IT buy-in. So supported in part by a College Access Challenge Grant from the Department of Education, we collaborated with other instructors to create online homework sets for three classes: Elementary Algebra, Intermediate Algebra, and Statistics for Behavioral Sciences I. After experimentation, some of these question pools are now created by Mathematica programs that can generate data sets from specified distributions, generate random polynomials that factor in a given way, create image files of histograms, scatterplots, and so forth. These programs produce files that can be read by the software package, Respondus, which then uploads the questions into Blackboard Learn, the course management system used by the Connecticut State University system. Finally, we summarize five classes worth of student performance data along with lessons learned while working on this project.

## 1. Introduction

We became interested in developing online homework for students in the introductory mathematics and statistics classes from different perspectives. The first author was teaching and tracking students in developmental mathematics in order to improve student outcomes, supported in part by a College Access Challenge Grant. This includes outreach to faculty and students at high schools geographically near Central Connecticut State University (CCSU), creating summer programs such as Bridges to help incoming freshman succeed in our developmental classes, and assessing the progress of these students from semester to semester. Consequently, she needed to have online questions that could be shared with high schools, could be used across classes that used different textbooks, and could be made accessible to students between semesters or over the summer even if they were not taking a math class. Resources that could be used in all these ways precludes using, for example, an online homework platform provided by a textbook publisher, so creating a problem database that was unrestricted by outside interests became important.

In his introductory statistics classes, the second author became dissatisfied with the traditional model of teaching from a textbook by just lecturing and wanted to incorporate a more data driven approach, such as the one proposed by Cobb (1991), one that involved collecting data in class to answer questions of interest to students. Eventually he replaced the textbook with PowerPoint slides and activities, but wanted online exercises that would give feedback to students on their ability to do key tasks that are needed when one is analyzing a data set such as giving a rough estimate of the correlation given a scatterplot. However, by not using a textbook, a publisher's homework platform was no longer available even if it were desired, so creating a problem database again was a solution. Eventually, we realized that we were both writing online problems, and we decided to pool our efforts and areas of expertise.

Even with two people, the task of creating questions by hand was time consuming and daunting. Moreover, because students can and do work together, it is best to have online homework where each student gets mathematically equivalent question that only varies by the values used. Course management systems vary, but all allow question pools, but this requires tedious cutting and pasting if done by hand. Some systems allow question templates and can generate random values, but Blackboard Learn, which is used by CCSU, does not have this feature.

One solution to this problem is to use an open access online homework bank such as WeBWorK, which is supported by the Mathematical Association of America and is described in Gage et al. (2002). This is an appealing option, but one that requires buy-in from the university's administration and the Information Technology Department, which has not happened yet.

Another solution to this problem is to automate the process, which is the one we pursued. This paper outlines how Mathematica has been used to generate question pools in a format that is easily uploaded to Blackboard Learn via a program called Respondus. Although the authors' experiences are with these specific software packages, course management systems generally allow a person to create a text or Word file that can be uploaded, so the basic idea is more widely applicable.

## 2. Online Homework

The effectiveness of online homework has been often studied and debated. The literature is mixed on whether or not it is more effective than the traditional paper homework that is graded by hand, which, for example, is discussed in Burch and Kuo (2010) and Zerr (2007). However, from direct experience in teaching mathematics and statistics classes, we know that online homework is better than (1) not assigning any or (2) assigning it but not grading it. Moreover, students like getting immediate feedback, and when given a choice to do online homework or not, they often pick the former as will be shown below.

We see online homework as a supplementary tool for assessment of basic skills and computation. In introductory mathematics and statistics courses, some students struggle with basic tasks like solving a linear equation or computing a z-score. Such deficiencies are important to discover quickly, which can easily be done with online homework because it is easy to create questions that test exactly these types of specific, simple skills. Using multiple choice or fill-in-the-blank questions is limiting in general, but they are well suited for testing the lower levels of Bloom's taxonomy for the cognitive domain such as the subject knowledge and comprehension levels as described in Bloom et al. (1956). This allows the teacher to spend time on assessing classroom activities, doing group projects, creating deeper exams that require thought and insight, and so forth.

These latter tasks require the higher levels of the taxonomy: application, analysis, evaluation, and synthesis.

Although creating our own pool of questions is an ongoing task that has required a fair amount of work up to this point, we are already seeing many payoffs. Our questions are customized exactly the way we want and are independent of any textbooks. Students can access these questions at no cost and whenever they desire to do. So they can study over school breaks, or be high school students thinking about coming to CCSU, or use these for review for higher-level courses. In particular, this ability to do community outreach and embedded remediation would not be possible in the traditional publisher's homework platforms, and as we collect outcome data, we believe that this will help students come to CCSU better prepared and get through their degree programs quicker. Now that our reasons for wanting online homework have been outlined, the next sections focus on some of the specific choices that have been made in this project.

## 2.1 Answer Formats

Different course management systems allow different types of questions. However, all systems have at least these two options: multiple choice (MC) and fill in the blank (FITB), which are the types we have used. Both of these have pluses and minuses, but knowing the potential pitfalls and advantages allows one to use these well.

### *2.1.1 Multiple Choice*

There is no doubt that MC has disadvantages. First, creating good distractor answers requires teaching experience and insight into student misunderstandings of the course material. Second, as noted above, MC applies best only to the lower levels of Bloom's taxonomy: subject knowledge and comprehension. Third, MC allows guessing. However, there is a fourth and more troubling drawback. As pointed out in Section 1.1 of Sangwin (2013), for some types of problems, checking potential answers is easier than solving the original problem. For example, solving $x^2 - 5x + 6 = 0$ by either factoring or the quadratic formula is harder than substituting $x = 2$ or $x = 3$ into the equation. However, checking answers is not always easier than a direct solution. For instance, solving "what is the probability that the sum of two dice is 5" requires knowing how to count up the possibilities, but if the answers are given in decimal form, that does not give a student an easy short cut.

Moreover, MC does have advantages. First, it is easy to enter such questions into a computer. Second, compared to FITB, there is no ambiguity in the student's answer. Finally, MC is quite good for checking student misunderstandings, especially "buggy rules," a term used in Section 6.10 of Sangwin (2013). For example, some students do think that $(a + b)^2 = a^2 + b^2$, which makes the right hand side a good distractor.

Given the above pros and cons, MC is useful for detecting common mistakes that students make even though it can be limiting. Fortunately, some of its shortcomings can be bypassed by using FITB, as discussed below.

### *2.1.2 Fill in the Blank*

FITB is not panacea because it has one big disadvantage: there are multiple ways to write a correct solution. For example, what is ½ - ¼ - ¼? Of course it is 0, but this could be written 0/4, -0/4, 0/2, 0, 0., 0.0, etc. Including instructions on what is a permissible answer such as "write your solution to two decimal places" reduces the possibilities, but 0.00 and .00 both satisfy these instructions, and at some point, long detailed instructions get cumbersome. Moreover, simple math problems can have solutions in multiple forms, each of which is defensible. For instance, "expand $(b + a)^2$" could be answered in the following ways: $a^2 + 2ab + b^2$, $b^2 + 2ab + a^2$, $b^2 + 2ba + a^2$, $b^2 + a^2 + 2ba$, $a^2 + b^2 + 2ab$, etc. One could try to explain to introductory-level students how to define lexicographic order on monomials to determine which one of these possibilities is "correct," but that would be confusing, not helpful. In practice, both specifying the form of the answer and having the online system give full credit to several possible forms can resolve this problem. An example of this is discussed below.

Of course, FITB does have advantages, which are easy to state. First, unlike MC, guessing by students is no longer effective. Second, students are more likely to realize when they need help. For our purposes of probing our students' basic subject knowledge and comprehension, both MC and FITB are valuable, and we have not tried using other forms of questions supported by Blackboard Learn. Moreover, by automating the writing of homework questions, we have been able to build up large question pools relatively quickly, the details of which are discussed in the next section.

## 3. Automating Question Creation

Making questions by hand is tedious and the chance of making at least one error is high when one makes a question bank. However, with the availability of computer algebra systems, it is not hard to write a program that both generates a question along with its answer. Moreover, for a MC question, distractors can be made using buggy rules, and for FITB, alternate valid forms of the correct answer can be generated. This project uses Mathematica to create HTML versions of the questions, answers, and if needed, graphics. Microsoft Word accepts HTML as input, which then can be saved as a Word document. This, in turn, is read into the package, Respondus, (see https://www.respondus.com/), which can upload questions to a variety of course management systems, including Blackboard Learn. For ease of programming and keeping track of which types of questions have been created, each question template corresponds to exactly one Mathematica function.

### 3.1 MC Example: Estimating the Correlation from a Scatterplot
Here is an example where MC is the best option. Given a scatterplot with, say, a hundred points, no one can make an estimate accurate to two decimal places, and even one decimal place is challenging. But recognizing the sign of the correlation (positive for upward trending values, negative for downward trending) and being able to distinguish between $r = 0.8$ vs. $r = 0.2$ is useful for the students to learn.

To start the process, the Mathematica program generates the absolute value of a correlation from a Uniform[0.5, 0.8] distribution, and its sign is determined by a coin flip. Next, a sample size is picked from a Uniform[50, 200] distribution, then a bivariate normal random sample is generated with this correlation and sample size. The actual correlation is computed, which is the correct answer. The distractors are the negative of the correct correlation, and three additional values close to 1, 0, and -1, respectively.

One nice feature is the ability to create a graphic, save this as an image file, then create HTML that inserts this image into the question. Specifically, the scatterplot is saved as a PNG file by the `Export[]` function to a filename generated by the program such as "correlation0029.png," where the number is needed because a pool of questions is being created. The HTML code for including this image is ``, which is quite short. Code Sample 1 has the specific code for these two actions, where `data = {x, y}` is the simulated bivariate normal values. Figure 1 has the HTML code generated, and Figure 2 shows how this looks in Blackboard Learn, which is very similar to how it appears in Microsoft Word.

```
out = ListPlot[data, AxesLabel -> {"Variable 1", "Variable 2"},
   PlotRange -> {{0, Max[x, y]}, {0, Max[x, y]}}, ImageSize -> 500];
filename = "correlation" <> ToString[questionNumber] <> ".png";
Export[filename, out];

...

WriteString[strexam, "\n<img src=\"" <> filename <> "\">\n<BR><BR>"];
```

**Code Sample 1:** Mathematica code that (1) creates a scatterplot and exports it as a PNG image file, and (2) prints a tag to include this image in the HTML file it is writing.

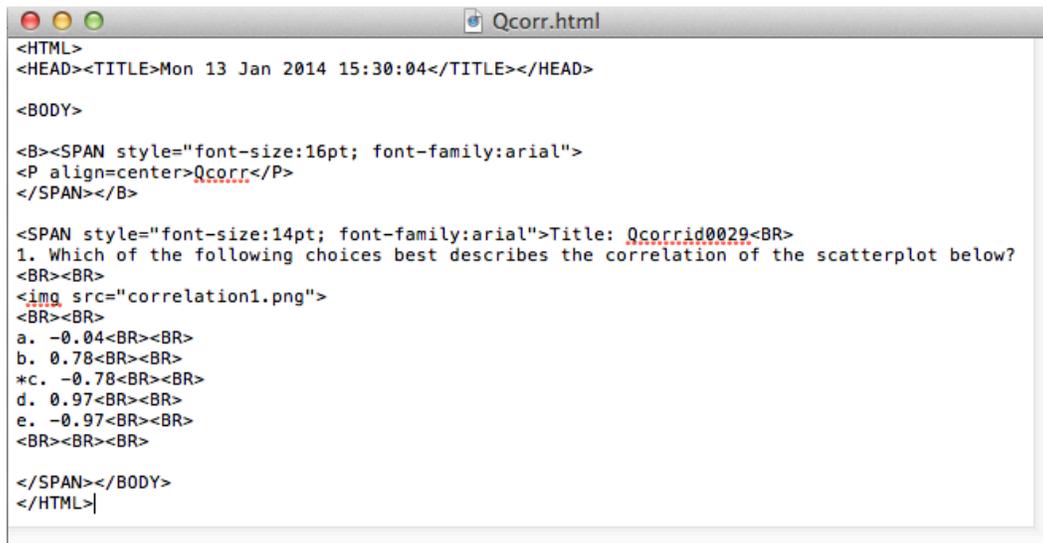

**Figure 1:** HTML code created by the Mathematica program. This can be directly opened by Microsoft Word, the results of which is shown in Figure 2.

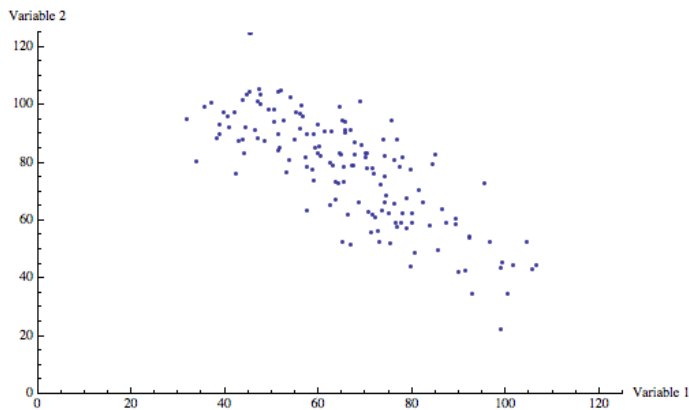

**Figure 2:** A MC question ready to be uploaded by Respondus into Blackboard Learn. The asterisk in front of answer c indicates that this is the correct choice.

### 3.2 MC Example: Area of a Trapezoid

Problems involving geometry obviously benefit from a graphic. This example, computing the area of a trapezoid given a plot of its sides and vertices, is one of a pair where the other is computing the perimeter. The image was made by combining two plots: one of the point grid and the other of the blue lines of the trapezoid itself. The MC possibilities for both types of question give the correct area and perimeter along with an incorrect area and perimeter. In addition, this example has answers that include a square root sign, which requires special formatting to look like what students are used to. However, Mathematica has a function called `TraditionalForm[]`, which produces typeset mathematics as an image, which can be used as MC options. Figure 3 shows an example of the HTML form of such a question, and Figure 4 shows how this appears.

One might argue that this type of question should be put into a FITB format. However, there is no need to decide between formats because both could be used. For example, a homework set could have the MC version first then the FITB. Although there are sophisticated adaptive systems like ALEKS that use artificial intelligence to determine optimal question sequencing, with thought a person with teaching experience can make an ordering that probes the strengths and weaknesses of a student, too.

```
<HTML>
<HEAD><TITLE>Sun 11 May 2014 22:28:14</TITLE></HEAD>

<BODY>

<SPAN style="font-size:14pt; font-family:arial">
Title: trapezoid-1<BR>
1. Which of the following choices is the area of the figure shown below?<BR><BR>

<BR><BR>
a. <BR>
b. <BR>
c. <BR>
*d. <BR>
<BR><BR>

</SPAN></BODY>
</HTML>
```

**Figure 3:** HTML code for the question displayed in Figure 4. Note that the four multiple choice answers are all images created by Mathematica's `TraditionalForm[ ]`.

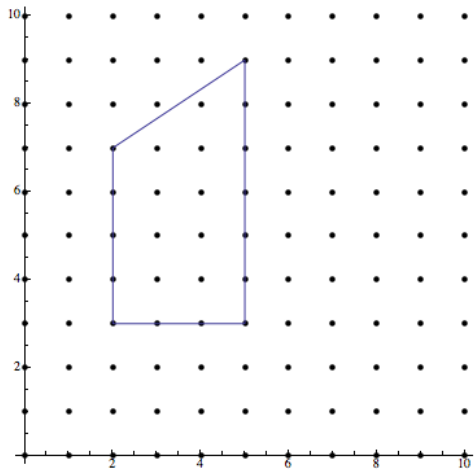

Title: trapezoid-1
1. Which of the following choices is the area of the figure shown below?

*a. 15
b. $10 + \sqrt{13}$
c. $13 + \sqrt{13}$
d. 18

**Figure 4:** A MC question asking for the area of the trapezoid. Choice c is the perimeter, b is a perimeter distractor, and d is an area distractor.

### 3.3 FITB Example: Solving Linear Equations

Solving a linear equation is a key skill in remedial mathematics and introductory statistics. For example, z-scores and regression lines are both linear. For students whose math skills are poor, having a sequence of questions is valuable. For example, one can start a homework set with a linear equation with integer coefficients and an integer answer. Next could be an equation with integer coefficients with a rational answer, and eventually a student needs to be comfortable with decimal coefficients, which will generally have a decimal answer. For example, a regression line for real data generally would have decimal coefficients.

Figure 5 gives an example of a linear equation with both integer coefficients and answer. This can be obtained by (1) picking the solution first, then (2) solving for one of the constant coefficients. For example, suppose $x = 5$ is the solution, then solving the equation below for $c$ after substituting this value for $x$ gives the desired type of problem. Here $c = 8$, which is an integer as promised.

$$3x - 2 = x + c$$

Although these three problems were forced to give an integer answer, the student is not told this, so he or she may not enter the answer as an integer without a decimal point. Blackboard Learn does not have an option to compare student input with a regular expression, but it is not hard to list a few alternative answers that a student might enter. For example, question 2 in Figure 5 gives full credit for -4, -4., -4.0, and -4.00. It is true that -4.00000 is also correct, but in the rare case a student entered such a value, he or she is likely to ask why credit was not given and would start to use less trailing zeros.

```
                              LinEqIntCffIntSolTXT.txt
Type: FMB
Title: LinEqIntCffIntSol-0031
1.  Solve for the value of z that makes the following equation true.

17z + 842 = -5z - 16

z = [-39, -39., -39.0, -39.00]

Type: FMB
Title: LinEqIntCffIntSol-0032
2.  Solve for the value of z that makes the following equation true.

-2 - 13z = 10z + 90

z = [-4, -4., -4.0, -4.00]

Type: FMB
Title: LinEqIntCffIntSol-0033
3.  Solve for the value of y that makes the following equation true.

-18y + 1066 = 16y + 8 + 12y

y = [23, 23., 23.0, 23.00]
```

**Figure 5:** Three linear equations with both integer coefficients and an integer solution.

In developmental mathematics, working with rational numbers is important, so having students solve linear equations with rational coefficients and requiring the answer to be a rational number is useful. Here getting rid of all the denominators by multiplying through by an integer is key. The easiest way to do this would be to use the least common multiple of all the denominators, but any common multiple would suffice. Examples of this type of problem are given in Figure 6. For problem 3, not that the least common multiple of 6 and 4 is 12, but a student might also use 6*4 = 24 to rationalize the fractional coefficients, hence there are two answers that are accepted: 9/14 and 18/28. Finally, Figure 7 shows how the first question in Figure 6 appears in Blackboard Learn.

```
                    LinEqRatCffRatSolTXT.txt
Type: FMB
Title: LinEqRatCffRatAns-0031
 1.  Solve for the value of x that makes the following equation true.
Enter your answer as a rational number.

    2x/3 = 3/4

    x = [9/8]

Type: FMB
Title: LinEqRatCffRatAns-0032
 2.  Solve for the value of y that makes the following equation true.
Enter your answer as a rational number.

    4y/3 = -5/3

    y = [-5/4, -15/12]

Type: FMB
Title: LinEqRatCffRatAns-0033
 3.  Solve for the value of e that makes the following equation true.
Enter your answer as a rational number.

    7e/6 = 3/4

    e = [9/14, 18/28]
```

**Figure 6:** Three linear equations with rational coefficients that require a rational answer from the student.

**Question 1**

Solve for the value of x that makes the following equation true. Enter your answer as a rational number.

2x/3 = 3/4

x = ☐

**Figure 7:** How Question 1 in Figure 6 appears in Blackboard Learn.

The above examples give an idea of the types of questions we have created so far. But unless students put effort into doing the online homework, it will not have much impact. Up to this point we have five classes worth of data, which we describe in the next section.

## 4. Student Responses and Assessment

In the two sections of Intermediate Algebra shown in Figure 8, there is a positive correlation between the online homework grades for the semester and the final course grade. Because the former is used to compute the latter, an upper trend is not surprising, but there is clearly much variability about this regression line. However, there are no points below the red line, which means that no students had a high homework score and a low final grade, which is encouraging.

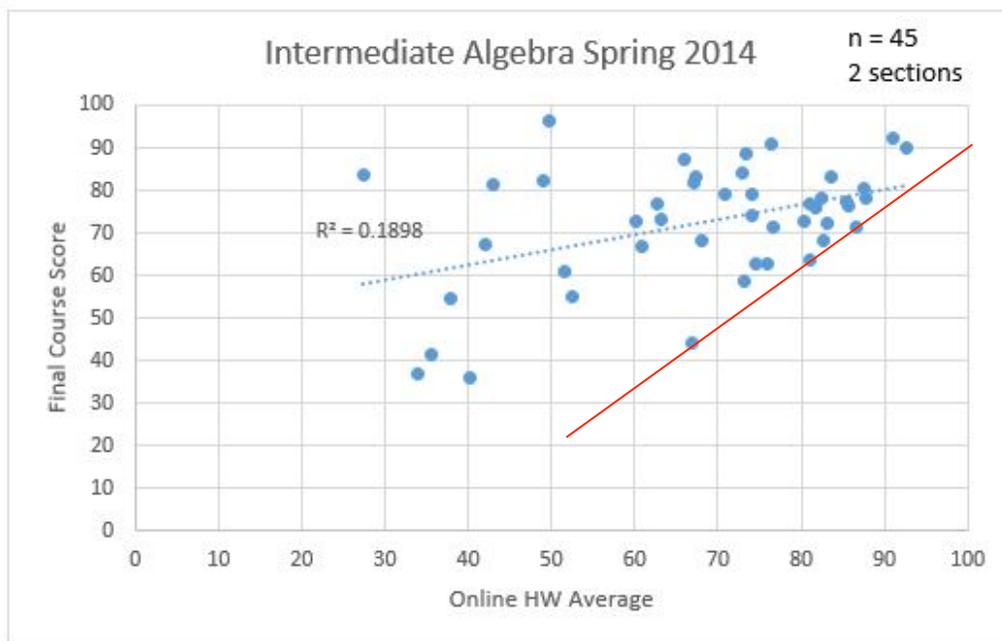

**Figure 8:** This scatterplot compares the course grade to the online homework grade for two sections of Intermediate Algebra. The data shows an upward trend, and note that there are no students below the red line.

In Introductory Statistics, the course grade was the average of the best four of the following six items: exams 1, 2, 3, and 4, the final, and all the online homework combined converted to a percentage. Consequently, a student could skip the homework completely, but Figure 9 shows that only two students actually did this, and only two more stopped doing the homework early in the semester. Moreover, these four data values are influencing the regression line such that its slope is lower than it would be without these. Redoing the regression without these four students, $R^2$ is 0.6292, so the upward trend is quite strong.

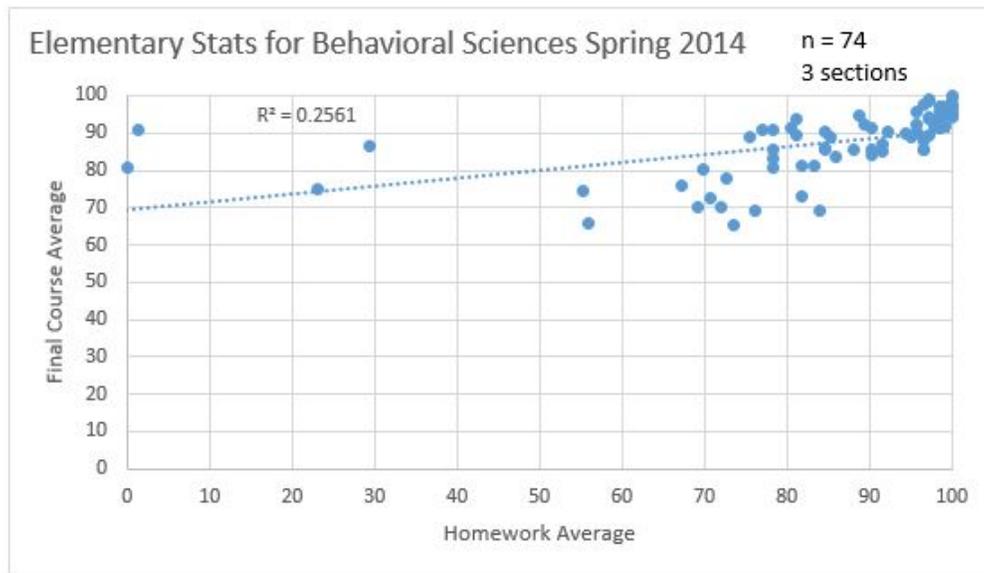

**Figure 9:** This is the same type of plot as Figure 8 for the three sections of Introductory Statistics. The four values on left side look influential on the regression and represent students who stop doing online homework early in the semester.

We are still collecting and plan to track students as they progress through CCSU, but our initial findings are encouraging: the homework is positively correlated with the course grade, and in Introductory Statistics where students are allowed to skip the homework, they overwhelmingly choose not to do so. Creating enough problems for an entire semester is a demanding task, and we hope one day to join our efforts with one of the open source question pools such as WeBWorK. However, given our current freedom from being locked into one publisher and the ability to provide unfettered access to both potential and current students, we are happy with our initial decision to create our own online homework.

## Acknowledgements

Several people have helped and encouraged us in our efforts. Most of all, we thank David Oyanadel and Jennifer Nicoletti from The Instructional Design and Technology Resource Center (IDTRC) along with their staff.